\documentclass[aps,twocolumn,nofootinbib,superscriptaddress,preprintnumbers,pra,10pt]{revtex4-1}

\usepackage{float}
\usepackage{arydshln}
\usepackage{colortbl}
\usepackage{amsmath,amssymb}
\usepackage{dsfont} 
\usepackage{hyperref}
\usepackage{graphicx}
\usepackage{enumitem}
\usepackage{arydshln}
\usepackage{mathtools}
\usepackage{bbold}
\usepackage{soul}
\usepackage{slashed}
\usepackage{amsmath,bm}
\usepackage{yfonts}
\usepackage{lipsum}
\usepackage{slashed}
\usepackage{mathtools}
\usepackage{physics}

\makeatletter
\g@addto@macro\bfseries{\boldmath}
\makeatother

\newcommand{\be} {\begin{equation}}
\newcommand{\ee} {\end{equation}}
\newcommand{\bea} {\begin{eqnarray}}
\newcommand{\eea} {\end{eqnarray}}
\newcommand{\no} {\nonumber}
\newcommand{\cO}{{\mathcal O}}
\newcommand{\cR}{{\mathcal R}}
\newcommand{\cC}{{\mathcal C}}
\newcommand{\cL}{{\mathcal L}}

\newcommand{\cB}{{\mathcal B}} 
 
\newcommand{\llpair}{\bar\ell\ell}
\newcommand{\mmpair}{\bar\mu\mu}
\renewcommand{\Re}{{\rm Re}}

\newcommand{\aria}[1]{{\color{green} #1}}

\begin{document}

\preprint{ZU-TH-36/22}
\title{Semi-inclusive $b\to s\llpair$ transitions at high $q^2$}
 
\author{Gino Isidori }
\author{Zachary Polonsky}
\author{Arianna Tinari}

\affiliation{Physik-Institut, Universit\"at Zu\"rich, CH-8057 Z\"urich, Switzerland}

\begin{abstract}
\vspace{5mm}
We present an updated Standard Model (SM) estimate of the inclusive $b\to s\llpair$ rate at high dilepton invariant mass ($q^2\geq 15~{\rm GeV}^2$).  We show that this estimate is in good agreement with the result obtained summing the SM predictions for the leading one-body modes ($K$ and $K^*$) and the subleading non-resonant $K\pi$ channel (for which we also present an updated estimate).  On the contrary, the semi-inclusive sum based on data exhibits a deficit compared to the inclusive SM prediction in the muon modes. The statistical significance of this deficit does not exceed $2\sigma$, but is free from uncertainties on hadronic form factors, and fully compatible with the deficit observed at low-$q^2$ on the exclusive modes. The implications of these results in conjunction with other SM tests on $b\to s\mmpair$ modes are briefly discussed.

\vspace{3mm}
\end{abstract}

\maketitle
\allowdisplaybreaks

\section{Introduction}

The ultimate goal of studying  $b\to s\llpair$ decays is to probe the short-distance structure of the corresponding flavor-changing neutral-current (FCNC) 
  amplitudes. By doing so we perform precise tests of the Standard Model (SM) probing, at the same time, motivated beyond-the-SM (BSM) theories. The presence of narrow charmonium resonances poses challenges in extracting short-distance information for both exclusive and inclusive $b\to s\llpair$ decays if the invariant mass of the dilepton pair,
$q^2 = (p_{\bar{\ell}} + p_\ell)^2$, is close to the resonance masses.   
This is why precise SM tests are confined to  $q^2 \lesssim 6-8\,$GeV$^2$ (low-$q^2$ region) and $q^2 \gtrsim 14-15\,$GeV$^2$ (high-$q^2$ region).  It is important to study both these regions as they are sensitive to different short-distance physics and, most importantly, they experience a different interplay between short- and long-distance dynamics. For a similar reason, it is important to study $b\to s\llpair$ transitions both at the exclusive and inclusive levels. 

In the last few years measurements of rates and angular distributions of the exclusive $B \to K^{(*)}\mmpair$ decays by LHCb~\cite{LHCb:2013ghj,LHCb:2014cxe,LHCb:2015svh}  have shown significant tensions with the corresponding SM predictions, especially in the low-$q^2$ region  (see e.g.~Ref.~\cite{Gubernari:2022hxn,Gubernari:2020eft,Alguero:2023jeh,Alguero:2018nvb,Altmannshofer:2021qrr,Hurth:2020ehu} for recent analyses). All the attempts to compute the decay amplitudes from QCD agree on the observed tension. However, using a more agnostic data-driven approach, some doubts about the reliability of the theory errors have been raised in Ref.~\cite{Ciuchini:2019usw,Ciuchini:2022wbq}.
The goal of this paper is to attempt to shed light on this issue by looking at the inclusive $B$-meson decay rate, $\Gamma(B\to X_s \llpair)$, in the high-$q^2$ region. 
This observable provides complementary information on  $b\to s\llpair$ amplitudes, being 
affected by qualitatively different uncertainties with respect to those appearing in the exclusive modes in the low-$q^2$ region.

The heavy-quark expansion in the high-$q^2$ region is an expansion in $O(\Lambda_{\text{QCD}}/(m_b - \sqrt{q^2}))$~\cite{Buchalla:1998mt},
which  converges less rapidly with respect to the  $O(\Lambda_{\text{QCD}}/m_b)$
expansion at work in the low-$q^2$ region (see Ref.~\cite{Huber:2005ig,Huber:2019iqf,Huber:2020vup}). However, as pointed out by Ligeti and Tackmann~\cite{Ligeti:2007sn}, non-perturbative uncertainties in the high-$q^2$ region can be greatly reduced by computing the ratio of the FCNC transition and the
$b\to u$ charged-current decay,
\be
 R^{(\ell)}_{\rm incl}(q_0^2) = \frac{\displaystyle\int_{q^2_0}^{m_B^2} d q^2 \frac{d \Gamma(B \to X_s \llpair)}{d q^2} }{
 \displaystyle\int_{q^2_0}^{m_B^2} d q^2 \frac{d \Gamma(B \to X_u \bar\ell\nu)}{d q^2} }\,,
 \label{eq:R0}
\ee
where $q_0^2$ is the lower cut on $q^2$.
The hadronic structure of the two transitions is very similar
($b\to q_{\rm light}$ left-handed current), leading to a significant cancellation of  non-perturbative uncertainties when
taking the ratio on an approximately equal phase space. Thanks to the recent experimental measurement of the 
$B \to X_u \bar\ell \nu$ inclusive rate as function of $q^2$
by Belle~\cite{Belle:2021ymg}, the procedure proposed in~\cite{Ligeti:2007sn} of computing  
the ratio~(\ref{eq:R0}) to predict $\Gamma(B \to X_s \llpair)$ can finally be put in place. 

On the experimental side, the $\Gamma(B \to X_s \llpair)$ rate at high-$q^2$ is not fully available. However, in this kinematic region, only a few decay modes are relevant and we can replace the inclusive sum with the sum over a limited set of exclusive modes. To this end, we update the prediction for the non-resonant 
$B\to K\pi$ mode at high $q^2$ presented in Ref.~\cite{Buchalla:1998mt}.
By doing so, we show that for $q_0^2 = 15~{\rm GeV}^2$ the inclusive rate is largely dominated by the two leading one-body modes ($B\to K$ and $B\to K^*$), with 
$B\to K\pi$ representing an $O(10\%)$ correction and additional multi-body modes
being further suppressed. We also show that the semi-inclusive rate obtained by summing the SM predictions for the leading one-body modes and the $B\to K\pi$ channel is in good agreement with the fully inclusive SM prediction obtained by means of $ R^{(\ell)}_{\rm incl}(q_0^2)$ in Eq.~(\ref{eq:R0}).
In other words, on the one hand we cross-check the SM inclusive prediction, on the other we validate the procedure to extract the (semi-)inclusive rate from data (while waiting for a fully inclusive measurement).  
As we shall show, the analysis of present data confirms the tension between data and SM prediction in the muon modes.

The paper is organized as follows: in Section~\ref{sec:Lagrangian} we review the $b\to s\llpair$ effective Lagrangian, pointing out the usefulness of a change of basis for the FCNC operators compared to the standard choice. In Section~\ref{sect:II} we present an updated estimate of 
\be 
\Gamma(B \to X_s \llpair)_{[15]} \equiv \Gamma(B \to X_s \llpair, q^2 \geq 15~{\rm GeV^2})\,,
\ee 
by means of (\ref{eq:R0}), following the analysis of Ref.~\cite{Ligeti:2007sn}. 
In Section~\ref{sect:III} we present the updated estimate of the $B\to K\pi\llpair$ rate for $q^2 \geq 15~{\rm GeV^2}$.
In Section~\ref{sect:IV} we compare inclusive vs.~semi-inclusive predictions within the SM, and the inclusive SM rate vs.~data (semi-inclusive, muon modes only). Finally, in Section~\ref{sect:V} we discuss the implications for the Wilson coefficients, encoding short-distance physics, inferred by the comparison with data. 
The results are summarised in the Conclusions.

\section{The $b\to s\llpair$ effective Lagrangian}\label{sec:Lagrangian}

The effective Lagrangian valid below the electroweak scale relevant to $b\to s\llpair$ transitions 
is conventionally written as
\bea
	\cL^{b\to s \llpair}_{\rm eff} &=& \frac{4G_F}{\sqrt{2}}  \frac{\alpha_e}{4\pi}\left(V^*_{ts}V_{tb}\sum_i C_i \cO_i + \text{h.c.} \right) \no\\
 				&&+\ 	\cL^{N_f =5}_{\rm QCD\times  QED} \,,
	\label{eq:bsll}
\eea
where we have used CKM unitarity, and neglected the tiny $O(V^*_{us}V_{ub})$ terms, 
to normalize all the flavor-changing operators in terms of a single CKM coefficient.

The only $\cO_i$ with $b\to s\llpair$  matrix elements which are non-vanishing at 
tree level are the electric-dipole operator 
\be
	\cO_7 =\frac{m_b}{e}(\overline{s}_L\sigma_{\mu\nu} b_R)F^{\mu\nu}\,,  
\ee
and the two FCNC semileptonic operators	
\be
\cO_{9}=(\overline{s}_L\gamma_{\mu} b_L)(\overline{\ell} \gamma^{\mu}\ell)\,, \quad
\cO_{10}=(\overline{s}_L\gamma_{\mu} b_L)(\overline{\ell} \gamma^{\mu}\gamma_5\ell)\,.		
\ee
For reasons that will be clear in the following, we find it convenient to perform a change 
of basis $\{ \cO_{9},  \cO_{10} \} \to \{ \cO_V,  \cO_L \}$, where 
\be 
	\!\cO_V =(\overline{s}_L\gamma_{\mu} b_L)(\overline{\ell} \gamma^{\mu}\ell)\,, \quad
	\cO_L =(\overline{s}_L\gamma_{\mu} b_L)(\overline{\ell}_L \gamma^{\mu}\ell_L)\,,
 \label{eq:OVOL}
\ee
such that 
\be
 C_V = C_9 + C_{10}\,, \qquad  C_L = -2 C_{10}\,.
\ee
The new basis allows us to separate effective interactions which originate
by different underlying dynamics and behave differently in the evolution 
from high scales ($\mu_0 \sim m_t$) down to low scales ($\mu_b \sim m_b$).
To better understand the different structures of these two operators, it is worth looking 
at  the corresponding Wilson coefficients 
at the lowest non-trivial order. 

\subsection{The $\cO_L$ operator}\label{sec:OL}

The purely left-handed operator is completely dominated by short-distance dynamics: it is generated at high scales by the top-quark Yukawa and $SU(2)_L$ interactions and, to a large extent,  
it does not evolve in the effective theory or mix with any other effective operator.

In the case of 
$\cO_L$, the presence of ${\alpha_e}$ in the normalization of $\cL^{b\to s \llpair}_{\rm eff}$ 
is rather misleading: this is evident if we look at the overall coefficient of 
$\cO_L$  (modulo the CKM factor), namely~\cite{Bobeth:2013tba} 
\be
\cC_{L} =  \frac{4G_F}{\sqrt{2}}  \frac{\alpha_e}{4\pi} C_L\,.
\ee
The corresponding one-loop expression is
\be
\cC_{L}^{\rm (0)} =  \frac{  2 G_F^2 m_W^2}{\pi^2}  Y_0(x_t)  =   \frac{y_t^2}{16 \pi^2 v^2}\left[ 1 + O(g^2/y_t^2)\right]\,,
\label{eq:CL0}
\ee
where $Y_0(x_t=m_t^2/m_W^2)\approx 0.98$ is the (finite) 
one-loop function~\cite{Inami:1980fz}, defined as
in~\cite{Buchalla:1995vs}.
As can be seen, the expression of $\cC_{L}^{\rm (0)}$ 
depends only on the top-quark Yukawa coupling ($y_t$) and the $SU(2)_L$ coupling 
($g$), once we normalize the effective interaction via the Higgs 
vacuum expectation value $v = (2 \sqrt{2} G_F)^{1/2} \approx 246$~GeV.
Moreover, this coefficient is non-zero in the so-called gaugeless limit of the 
SM (i.e.~in the limit $g\to 0$ and $y_t\not=0$, see e.g.~\cite{Barbieri:2008zt}).

The separation of $\cO_L$ from all the other operators in $\cL^{b\to s \llpair}_{\rm eff}$ 
is guaranteed by the fact that, in the limit where we neglect light Yukawa couplings, the RG evolution arises only from QCD and QED, which are vector-like theories.
Note also that the bilinear quark current in $\cO_L$ is a conserved current,  hence there is no mixing and no contribution in the RG evolution of $C_L$ at any order in QCD. 
A small anomalous dimension and a tiny mixing with other operators arise only from higher-order QED corrections. 

The stability of $\cC_L$ under quantum corrections 
 is reflected by the small numerical difference between the leading (one-loop) result,
$\cC_{L}^{\rm (0)}\approx 1.7 \times 10^{-7}~{\rm GeV}^{-2}$, and the precise value estimated in~\cite{Bobeth:2013tba,Bobeth:2013uxa}, taking into account 
NNLO  QCD and EW corrections (which play an important role in reducing the scale uncertainty in the high-scale matching).
From the analysis of Ref.~\cite{Bobeth:2013uxa}, taking into 
account the updated input for $m_t$ (see Table~\ref{tab:inputsI})
we deduce
\bea
\cC_{L} (m_b) &=& (1.662 \pm 0.008) \times 10^{-7}~{\rm GeV}^{-2}\,,  \\ 
C_L (m_b)  &=&  8.38 \pm 0.04\,,
\label{eq:CLmub}
\eea
where in (\ref{eq:CLmub}) we have used the normalization (\ref{eq:bsll}) and, correspondingly,  the value of $\alpha_e(m_b)$ in Table~\ref{tab:inputsI}.

\begin{table}[t]
	\centering
	\begin{tabular}{|c|c|c|}
		\hline
		Parameter & Value & Reference \\
		\hline
		$m_t$  & 172.7(5)\ GeV & \cite{PDG}  \\
        $\alpha(m_b)^{-1}$ & 132.306(9) & \cite{PDG,Huber:2005ig}   \\
		$|V_{ub}|$ & $3.82(20)\times 10^{-3}$ & \cite{PDG}  \\
		$|V_{ts}|$ & $4.15(9)\times 10^{-2}$ & \cite{PDG} \\
        $|V_{tb}|$ & $0.990903(64)$ & \cite{PDG} \\
		\hline
	\end{tabular}
	\caption{Input parameters used for the computation of the inclusive rate.}
	\label{tab:inputsI}
\end{table}

\subsection{The $\cO_V$ operator}\label{sec:OV}

This effective operator receives contributions from all scales,
vanishes in the limit $\alpha_e\to 0$ (i.e.~in the limit ${s_W\to 0}$ at fixed $g$), and  it mixes with the four-quark operators already 
at the one-loop level.

The one-loop expression, obtained without resumming large logarithms, is 
\bea
C_V^{\rm (0)}(\mu) &=&  -4 Z_0(x_t) +\frac{4}{9} - \frac{4}{9} \ln\left(\frac{\mu^2}{M^2_W}\right) \no \\
&\approx& 0.22 -  0.89\times \ln\left(\frac{\mu^2}{m^2_b}\right)\,, 
\label{eq:CV0}
\eea
with $Z_0(x_t)\approx 0.67$ defined as in~\cite{Buchalla:1995vs}. 
The numerical result in (\ref{eq:CV0}) is only qualitative (given we have not 
resummed the large logarithms),  but it illustrates well the main features of $C_V$. There are two competing contributions that tend 
to cancel each other: i) the finite and largely scale-independent  
short-distance contribution, encoded in $Z_0(x_t)$, and 
ii) the charm-loop contribution generating large logarithms in the RG evolution from high scales to low scales. The cancellation becomes even more effective when the sizable QCD corrections are resummed via a proper treatment of the RG evolution. At NNLO accuracy~\cite{Bobeth:2003at}, 
using the numerical results in~\cite{Blake:2016olu},
we find
\be
C_V(\mu_b) = -0.01 \pm 0.14 \,,  \qquad  \mu_b\in[2,5]~{\rm GeV}\,,
\label{eq:CVmub}
\ee
with a significant residual (low) scale dependence, which is a remnant of the $\mu$ dependence in
Eq.~(\ref{eq:CV0}). 

\subsection{Four-quark operators}
\label{sec:charmloops}

Beside $\cO_L$ and $\cO_V$, an important role in 
 $b\to s\llpair$ amplitudes is played by the 
 four-quark  operators, whose matrix elements   
 are non-vanishing beyond tree level. 
 To high accuracy 
(i.e.~to first order in $\alpha_e$ and 
arbitrary order in $\alpha_s$), the contribution of four-quark 
operators can be expressed via a (process-dependent, non-local) modification of $C_V$:
\bea
&& C_{V,X_s}^{\rm eff} (q^2) = \no \\
&& \frac{\displaystyle{    \sum_{H_s \in X_s}
\left( C_V \langle H_s  \llpair  | \cO_V | B \rangle
+ \sum_i C_i 
 \langle H_s \llpair  |   \cO_i  | B \rangle  \right)  } }{ 
\displaystyle{ 
 \sum_{H_s \in X_s}
 \langle H_s   \llpair| \cO_V  | B \rangle  
} } \qquad 
\label{eq:CVeff}
\eea 
The sum over $H_s$ denotes the sum over all the hadronic states belonging 
to the final state $|X_s \rangle$.  Due to quark-hadron duality, we expect that 
for a sufficiently inclusive $|X_s \rangle$ the hadronic sum 
can be replaced by a partonic sum.

Evaluating the matrix elements of the $\cO_i$ in Eq.~(\ref{eq:CVeff})
in perturbation theory at lowest order in $\alpha_s$, leads to a process-independent expression that we denote 
$C^{\rm eff}_{V } (q^2)|_{\rm pert}^{(0)}$.
More precisely, the coefficient thus obtained does not depend on $|X_s \rangle$, provided this state has the valence-quark content of $\cO_V | B \rangle$. The expression of $C^{\rm eff}_{V  } (q^2)|_{\rm pert}^{(0)}$
 is the same for the fully inclusive mode or for an exclusive decay such as $B\to K\llpair$. Considering only the leading four-quark charm-quark operators\footnote{We define Wilson coefficients of four-quark operators as in~\cite{Blake:2016olu} (the definition of the operators differs by an overall factor  due to the different normalizations of 
$\cL^{b\to s \llpair}_{\rm eff}$).}
$\cO^c_{1,2}$, which have $O(1)$ Wilson coefficients,
one finds
\begin{equation}
C^{\rm eff}_{V  } (q^2) \big|_{\rm pert}^{(0)} \approx  C_V +  \Big(C^c_2 +\frac{4}{3}C^c_1\Big) \times  h(m^2_c,q^2) 
\label{eq:CVeff0}
\end{equation}
where $ h(m^2_c,q^2)$ is given in~\cite{Parrott_2023} and 
$h(0,q^2)= (4/9) \times \log\left(\mu_b^2/m_b^2\right)$.
Using the numerical expressions for the Wilson coefficients in~\cite{Blake:2016olu}
we find 
\bea
 {\rm Re}\left[ C^{\rm eff}_{V  } (q^2 = 15~{\rm GeV^2})\big|_{\rm pert}^{(0)} \right] &=& 0.43 \pm 0.26~,\label{eq:CVeff1}\\
 {\rm Re} \left[ C^{\rm eff}_{V } (q^2 = 1~{\rm GeV^2})\big|_{\rm pert}^{(0)} \right] &=& 0.13 \pm 0.13~.\quad 
  \label{eq:CVeff2}
\eea
The error, due to the scale dependence, is closely connected to the scale variation of $C_V(\mu_b)$ in (\ref{eq:CVmub}). 

Going beyond this approximation we can decompose $C_{V,X_s}^{\rm eff}$ for the inclusive case as
\be
C^{\rm eff}_{V,X_s } (q^2)  = 
C^{\rm eff}_{V,X_s } (q^2) \big|_{\rm pert} +  
C^{\rm eff}_{V,X_s } (q^2) \big|_{\rm n.p.}  
\label{eq:CVeffT}
\ee
The two terms on the r.h.s.~of 
Eq.~(\ref{eq:CVeffT})
denote the result 
obtained in perturbation theory,
considering partonic states, and 
possible additional non-perturbative contributions, respectively. 
The NLO corrections 
to the perturbative term, evaluated for the first time in~\cite{Ghinculov:2003qd} 
in the high-$q^2$ region,
are within the error band
 of the  leading contribution
 in Eq.~(\ref{eq:CVeff1}).

On general grounds, non-perturbative contributions are expected to be smaller 
than perturbative ones.
The latter start to lowest-order in $\alpha_s$ and are not power suppressed 
in the heavy-quark limit. The only notable exception is the $q^2$ region 
of the narrow charmonium
resonances, where large {\em local}
violations of quark-hadron duality 
do occur. We defer a more detailed 
discussion of 
possible non-perturbative contributions  to Section~\ref{sect:V}. In the following, 
we limit ourselves to consider perturbative contributions only. 

\section{Inclusive rate at high $q^2$}
\label{sect:II}

The comparison between the numerical value of $C_V$ in (\ref{eq:CVmub}) and $C_L$ in (\ref{eq:CLmub}) 
indicates that, within the SM, the local part of the $b\to s\llpair$ interaction 
has an approximate left-handed structure, 
as in the $b\to u\ell \bar\nu$ case.
In both processes we deal with a $b\to q_{\rm light}$ transition, 
hence non-perturbative effects in sufficiently inclusive distributions 
are expected to be very similar. 
In $b\to s\llpair$ transitions, 
corrections to a pure local left-handed interaction 
are generated by the matrix elements of $\cO_7$
and those of the 
four-quark operators (discussed in Section~\ref{sec:charmloops}). 
However, both these effects are quite small in the high-$q^2$ region. 
This is why the ratio (\ref{eq:R0}) 
provides a very interesting observable to perform precise SM tests, as pointed out first in Ref.~\cite{Ligeti:2007sn}.

In order to  compare this ratio with experiments,
it is important to define the treatment of
electromagnetic corrections below $m_b$.
As pointed out first in~\cite{Huber:2005ig},  
these give rise to $\log(m_\ell)$-enhanced terms 
in the $q^2$ spectrum which implies 
a sizeable suppression of the rate in the high-$q^2$ region and a corresponding enhancement at low $q^2$. The origin of this effect is the migration of events to low $q^2$ due to real photon emissions by the dilepton system. 
As shown in~\cite{Isidori:2020acz},
this effect is absent (and the electromagnetic corrections
related to scales below $m_b$ become tiny) if the cut employed to define the relevant kinematical region is  $q_0^2 = (p_B - p_X)^2$.   
The experiments whose data we use for comparison effectively utilize this definition when analyzing exclusive modes~\cite{Isidori:2022bzw}. We therefore do not include electromagnetic corrections (real or virtual) at scales below $m_b$ in our prediction of $\cB( B \to X_s \llpair)_{[15]}^{\rm SM}$.
Neglecting such corrections, the ratio (\ref{eq:R0}) 
becomes $m_\ell$-independent (within the SM) 
and we can therefore drop the 
corresponding lepton label.

To provide an updated numerical prediction of $ R_{\rm incl}(q_0^2)$ within the SM 
we re-express the result of Ref.~\cite{Ligeti:2007sn} in the $C_{L,V}$ basis, 
rather than in $C_{9,10}$ one. 
This way we can write 
\be
 R_{\rm incl}(q_0^2) = \frac{ |V_{tb} V^*_{ts}|^2 }{ |V_{ub}|^2 } \left[\cR_L + \Delta \cR_{[q_0^2]} \right]
 \label{eq:RinclCVCL}
\ee
where 
\be
\cR_L = \frac{\alpha_e^2 C_L^2 }{16 \pi^2}  = 
\frac{\cC^2_L}{ 8 G_F^2} 
\ee
and, for $q_0^2=15~{\rm GeV}^2$,
\bea
&& \Delta \cR_{[15]} = 
\frac{\alpha_e^2 }{8 \pi^2} 
\Big[ C_V^2 + C_V C_L  \qquad \no \\ 
&&\qquad +\,  0.485 C_L+ 0.97C_V +  0.93 +\Delta_{\rm n.p.} \no \\
&&\qquad +\,  C_7 (1.91 + 2.05 C_L + 4.27 C_7+4.1 C_V) \Big]\,.
\qquad
\eea
The $\cR_L$ term is the result obtained in the limit of purely left-handed interactions and 
identical hadronic distributions, while $\Delta \cR_{[q_0^2]}$ describes all the deviations from this ideal limit.
The numerical coefficients 
in $\Delta \cR_{[15]}$ take into 
account the (perturbative) matrix elements
of the four-quark operators,
integrated over $q^2$,
while $C_V$ is the 
($q^2$-independent) Wilson 
coefficient.
We denote by $\Delta_{\rm n.p.}$ the non-perturbative effects estimated in~\cite{Ligeti:2007sn}. The latter 
 do not include 
 $SU(3)$-breaking corrections due to light-quark 
masses.
A naive estimate of these effects, from the phase space differences on the leading hadronic modes, indicates corrections up to $8\%$ for $q_0^2=15~{\rm GeV^2}$. This is very similar in size to 
the error associated with $\Delta_{\rm n.p.}$ that, as we shall see,
does not represent the dominant source of uncertainty in the final estimate of 
$\cB( B \to X_s \llpair)_{[15]}^{\rm SM}$.
Similarly, possible non-factorizable contributions 
related to the broad charmonium resonances have not been 
explicitly included 
(we will  revisit this point in  Section~\ref{sect:V}).

The numerical expressions in the SM are
\bea
\cR_L^{\rm SM} &=&  (2.538 \pm 0.024)\times 10^{-5}\,,  \no \\
\Delta \cR^{\rm SM}_{[15]}   &=& (-0.03 \pm 0.14_{C_i} \pm 0.17_{\rm n.p.})\times 10^{-5}\,.
\label{eq:Rnum}
\eea
As it can be seen, $\Delta \cR^{\rm SM}_{[15]}$ is fully compatible with zero, but largely dominates 
the theoretical uncertainty in  (\ref{eq:R0}). The first error is due to the values of $C_V$ and $C_7$,\footnote{Since the scale variation  in Eq.~(\ref{eq:CVeff1})
is larger with respect to the one in Eq.~(\ref{eq:CVmub}),
we assume the former as \aria{a} conservative estimate 
of the scale uncertainty on $C_V$.}
while the second one is due to non-perturbative effects.

Using the numerical results in (\ref{eq:Rnum}), together with the experimental measurement of the 
$B \to X_u \bar\ell \nu$ inclusive rate for $q^2 \geq 15~{\rm GeV}^2$~\cite{Belle:2021ymg},\footnote{The result 
for $q^2 \geq 15~{\rm GeV}^2$ is obtained by means of the 
$q^2$ differential data in 
https://doi.org/10.17182/hepdata.131599}
\be
\cB( B \to X_u \bar\ell \nu)_{[15]}^{\rm exp} 
 = (1.50 \pm 0.24)\times 10^{-4}\,,
 \label{eq:BXuexp}
\ee
and the CKM inputs in Table~\ref{tab:inputsI},
we finally obtain
\bea
&& \cB( B \to X_s \llpair)_{[15]}^{\rm SM}  =  (4.5 \pm 1.0) \times  10^{-7}  
\label{eq:BRincl} \\
&& = 4.5 \times 10^{-7} \left[ 1 \pm  0.16_{\rm exp} \pm 0.11_{\rm CKM} \pm 0.09_{\rm \Delta \cR}  \right].\qquad
\label{eq:BRinclEr}
\eea
Note that the leading uncertainties are due to the experimental result in (\ref{eq:BXuexp}) and the CKM inputs. We could therefore expect a 
significant reduction of the total uncertainty in (\ref{eq:BRincl})
in the near future.

Our estimate of $\cR_{[15]}$
in (\ref{eq:Rnum}), and the corresponding 
SM prediction in (\ref{eq:BRinclEr}), 
are about $20\%$ higher with respect to 
the results obtained in Ref.~\cite{Huber:2020vup}
for the $\ell=e$ case. 
A large fraction of this difference can be attributed
to the different treatment of the real electromagnetic radiation.
The discrepancy indeed reduces 
to about $5\%$ when comparing with 
the results of Ref.~\cite{Huber:2020vup} 
in the absence of long-distance electromagnetic 
corrections.\footnote{See Table~5 in the
appendix of Ref.~\cite{Huber:2020vup} (note that  
the ratio $\cR$ defined in Ref.~\cite{Huber:2020vup} includes also 
the CKM factors).}

\section{The $B\to K \pi$ rate at high $q^2$}
\label{sect:III}

As anticipated, our goal is twofold. First, 
we cross-check the SM prediction in \eqref{eq:BRincl} with the corresponding semi-inclusive result, namely the sum of 
the SM predictions of the leading exclusive modes.
Second, we compare \eqref{eq:BRincl} with the experimental 
results in the high-$q^2$ region. A necessary ingredient
to achieve both goals is the SM prediction of the  $B\to K \pi$ rate, which we present in this section.


The  $B\to K \pi$ process must be treated with some care since 
it receives resonant contributions from ${\mathcal{B}(B\to(K^*\to K\pi)\llpair)}$, which are at least partially accounted for in the $B\to K^*\llpair$ branching fraction. To avoid double counting
these terms, we assume  $K^*$-dominance for the 
$p$-wave $B\to K\pi\llpair$ decay amplitude. 
In other words, we describe this part of the amplitude 
via the exchange of the $K^*$ resonance 
(also in the off-shell region, assuming a 
$q^2$-independent $K^*$ width). 
The $K^*$ resonance cannot contribute to the $s$-wave part of the amplitude, and the interference between $s$ and $p$ waves
cancels when integrated over the phase space 
at fixed $q^2$.  We therefore compute separately the $s$-wave component of the total branching fraction, $\mathcal{B}(B\to(K\pi)_s\llpair)$, that we treat as an independent decay channel.

At high $q^2$, the light mesons have very low recoil energies, $E_{\text{had}}\ll \Lambda_{\text{QCD}}$, and heavy hadron chiral perturbation theory (HHChPT) is valid in this region. We calculate the leading $s$-wave contribution to the total $B\to K\pi$ rate by computing the $B\to K\pi$ matrix element in HHChPT, and subtracting the corresponding $B\to K^*\to K\pi$ contribution, 
evaluated at the $K\pi$ threshold. The HHChPT calculation was performed in Ref.~\cite{Buchalla:1998mt}, and we independently verified the result. In order to simplify the comparison to the $B\to K^*\to K\pi$ matrix element obtained using the lattice results of Ref.~\cite{Horgan:2015vla}, we parameterize the matrix elements as
\begin{equation}\label{eq:KpiMEs}
    \begin{split}
        &\mel{K(p_K)\pi(p_\pi)}{\bar{s}\gamma^\mu(1 - \gamma_5)b}{B(p)} = \\ 
        &\hspace{0.5cm}-i\big(w_+ P^\mu + w_- Q^\mu + c\,q^\mu + i h \epsilon^{\mu\nu\rho\sigma}q_\nu P_\rho Q_\sigma\big)\,, \\
        &\frac{iq_\nu}{q^2}\mel{K(p_K)\pi(p_\pi)}{\bar{s}\sigma^{\nu\mu}(1 + \gamma_5)b}{B(p)} = \\ 
        &\hspace{0.5cm}-i\big(w_+' P^\mu + w_-' Q^\mu + c'\,q^\mu + i h' \epsilon^{\mu\nu\rho\sigma}q_\nu P_\rho Q_\sigma\big)\,,
    \end{split}
\end{equation}
where $P^\mu = p_K^\mu + p_\pi^\mu$ and $Q^\mu = p_K^\mu - p_\pi^\mu$. The relevant form factors are related to those in Ref.~\cite{Buchalla:1998mt}, by
\begin{equation}
    w_+^{(\prime)} = \frac{a^{(\prime)} + b^{(\prime)}}{2} + c^{(\prime)}, \quad w_-^{(\prime)} = \frac{b^{(\prime)} - a^{(\prime)}}{2} \,.
\end{equation}
Defining
\begin{equation}
    w_1 = m_B\Big(w_+ + \frac{d}{t}w_-\Big), \quad w_2 = m_B\,w_-\,,
\end{equation}
the leading term of the differential decay width in the expansion around the $K\pi$ threshold is given by
\begin{equation}\label{eq:HHDecay}
    \begin{split}
        \frac{d\Gamma}{ds_\ell} =& \frac{G_F^2 M_B^5}{192 \pi^3}|V_{ts}^* V_{tb}|^2\frac{\alpha^2}{4\pi^2}\frac{1}{32\pi^2}\frac{\pi}{4}\frac{\sqrt{t^2 x_1 x_2}}{(1 - t)^{3/2}} \\
        &\times \Bigg[\Big(\frac{1}{2}|C_L|^2 + |C_V|^2 + \Re(C_V^* C_L)\Big) F_9 \\
        & + 4 m_b^2|C_7|^2 F_7 + 4m_b\Re\Big\{C_7\Big(\frac{1}{2}C_L^* + C_V^*\Big)F_{97}\Big\}\Bigg]\\
        &\times (s^{K\pi}_\ell - s_\ell)^3 + O\big[(s^{K\pi}_\ell - s_\ell)^{7/2}\big]\,,
    \end{split}
\end{equation}
where the non-dimensional parameters are $x_1~=~m_K/m_B$, $x_2=m_\pi/m_B$, $t = x_1 + x_2$, $d = x_1 - x_2$, $s_\ell = q^2/m_B^2$, and $s^{K\pi}_\ell \approx 0.775$ is the value of $s_\ell$ at the $K\pi$ threshold.
The expression ($\ref{eq:HHDecay}$) does not take into 
account the matrix elements of the four quark operators.
The latter can easily be incorporated by the replacement 
$C_V \to C_{V,K\pi}^{\rm eff} ( s_\ell m_B^2)$.
Approximating $C_{V,K\pi}^{\rm eff}$ with 
$C_{V }^{\rm eff}|^{(0)}_{\rm pert}$ in 
(\ref{eq:CVeff0}), this replacement has 
a negligible numerical impact given the additional
sources of uncertainty.

The $F_i$ factors are given by
\begin{equation}
    \begin{split}
        F_7 =& |w_1'|^2 + \frac{4x_1 x_2}{t^2}(1 - t)|w_2'|^2\,, \\
        F_9 =& |w_1|^2 + \frac{4x_1 x_2}{t^2}(1 - t)|w_2|^2\,, \\
        F_{97} =& w_1'w_1^* + \frac{4x_1 x_2}{t^2}(1 - t)w_2'w_2^*\,.
    \end{split}
\end{equation}
The rate is largely dominated by the terms proportional to $w_+^{(\prime)}$
(i.e.~the form factors associated with the total hadron momentum in the matrix elements), which are the only ones relevant to the $s$-wave transition. At the $K\pi$ threshold, where we can still trust the HHChPT result,
we find
\begin{equation}
        w_+ = 79.46\,,  \qquad      w_+' = 16.49\,, 
\end{equation}
in units of GeV$^{-1}$. Conversely, evaluating the 
$B\to K^*\to K\pi$ contribution at the $K\pi$ threshold we find
\begin{equation} 
 \begin{split}
        w_+|_{\rm res} &= 10.23 + 1.05i\,, \\ 
        w_+'|_{\rm res} &= 2.47 + 0.25 i\,.
\end{split}
\end{equation}
Comparing these two results, we determine 
\begin{equation}
    \begin{split}
        w_+^s &= 69.23(62) - 1.05(6)i,\\[5pt]
        w_+^{\prime s} &= 14.02(11) -0.25(1) i\,,
    \end{split}
\end{equation}
again in units of GeV$^{-1}$. The uncertainties arise from both parametric inputs as well as lattice form factors. 

Using the $s$-wave form factors in Eq.~\eqref{eq:HHDecay}, and integrating for $q^2 \geq 15~{\rm GeV^2}$ we find\footnote{This value is obtained by summing the two isospin-related final states, and is equivalent to $3/2$ times the branching fraction featuring a charged pion in the final state.}
\begin{equation}\label{eq:Kpiresult}
    \mathcal{B}\big(B\to (K\pi)_s\llpair)^{\rm SM}_{[15]} = (5.8\pm2.5)\times 10^{-8}\,,
\end{equation}
where the errors are estimated from the fact that the NLO behavior in the expansion around the $K\pi$ threshold scales like $(s^{K\pi}_\ell - s_\ell)^{7/2}$. This error hugely dominates the parametric error, so the latter is not included in Eq.~\eqref{eq:Kpiresult}. Input parameters are given in Tabs~\ref{tab:inputsI} and \ref{tab:inputsEx}. Additionally, we use $g_\pi \sim 0.5$ for the HHChPT coupling constant and, along with $C_V$ and $C_L$ given in Sec.~\ref{sec:Lagrangian}, the remaining $C_i$ ($i=1,\dots,8$) are taken from Ref.~\cite{Blake:2016olu}.

\begin{table}[t]
	\centering
	\begin{tabular}{|c|c|c|}
		\hline
		Parameter & Value & Reference \\
		\hline
        $f_B$ & 0.1900(13)\ GeV & \cite{FLAG} \\
        $f_\pi$ & 0.13041(20)\ GeV & \cite{PDG14} \\
		$m_\pi$  & 0.137(3)\ GeV & \cite{PDG}  \\
        $m_K$ & 0.495(3)\ GeV & \cite{PDG}   \\
		$m_{B^\pm}$ & 5.27925(26)\ GeV & \cite{PDG}  \\
		$m_{{K^*}^\pm}$ & 0.89547(77)\ GeV & \cite{PDG} \\
        $\Gamma_{{K^*}^\pm}$ & 0.0462(13)\ GeV & \cite{PDG} \\
        $m_{{B^*}^\pm} - m_{B^\pm}$ & 0.04537(21)\ GeV& \cite{PDG} \\
        $m_{B_s} - m_{B}$ & 0.08742(24)\ GeV & \cite{PDG} \\
		\hline
	\end{tabular}
	\caption{Input parameters used for the computation of the exclusive branching fractions.}
	\label{tab:inputsEx}
\end{table}

\section{Inclusive  high-$q^2$ rate as sum of exclusive modes}
\label{sect:IV}

The exclusive $B\to K\llpair$ and $B\to K^*\llpair$ branching fractions can be computed using the form factors calculated in Refs.~\cite{Parrott:2022rgu,Horgan:2015vla}. Again, integrating for $q^2\ge 15$ GeV, we find
\begin{equation}\label{eq:leadingmodes}
    \begin{split}
        \mathcal{B}(B\to K\llpair)^{\rm SM}_{[15]} &= \big(1.31\pm 0.08_{\text{lat}}\pm 0.09_{\text{par}}\big)\times 10^{-7}\,\\
        \mathcal{B}(B\to K^*\llpair)^{\rm SM}_{[15]} &= \big(3.19\pm 0.21_{\text{lat}}\pm 0.22_{\text{par}}\big)\times 10^{-7}\,,
    \end{split}
\end{equation}
where ``lat'' refers to the uncertainty induced by the lattice form factors and ``par''   to the one from parametric inputs. 

In the above, we have used the narrow-width approximation to estimate 
the $K^*$ contribution. We can estimate the error due to a non-vanishing decay width by using the following double differential branching ratio, which is valid in the limit of constant width (or Breit-Wigner resonance):
\begin{equation}
    \begin{split}
        &\frac{d^2\mathcal{B}(B\to(K^*\to K\pi)\llpair)}{dq^2\,dp_{K\pi}^2} = \frac{d\mathcal{B}(B\to K^*\llpair)}{dq^2} \\[5pt]
        &\hspace{2cm}\times\frac{1}{\pi}\frac{m_{{K^*}}\Gamma_{{K^*}}\mathcal{B}(K^*\to K\pi)}{(p_{K\pi}^2 - m_{{K^*}}^2)^2 + m_{{K^*}}\Gamma_{{K^*}}} 
         \,.
    \end{split}
\end{equation}
Expanding in powers of $\Gamma_{K^*}/m_{K^*}$
and using $\mathcal{B}(K^*\to K\pi) \approx 1$~\cite{PDG}, gives
\begin{equation}
    \mathcal{B}(B\to K^*\llpair) \approx \Big(1 - \frac{1}{\pi}\frac{\Gamma_{K^*}}{m_{K^*}}\Big)[\mathcal{B}(B\to K^*\llpair)]_{\Gamma = 0}~.
\end{equation}
This implies an additional $O(1\%)$ correction that we can safely neglect due to the size of the form factor and parametric errors in Eq.~\eqref{eq:leadingmodes}.

The results in \eqref{eq:Kpiresult} and \eqref{eq:leadingmodes}
can be combined to define 
a ``correction factor'' from the two-body final state relative 
to the  one-body modes:
\begin{equation}\label{eq:correction}
    \Delta^{[15]}_{K\pi} = \frac{\mathcal{B}(B\to(K\pi)_s\llpair)_{[15]}}{\mathcal{B}(B\to K\llpair)_{[15]} + \mathcal{B}(B\to K^*\llpair)_{[15]}} = 0.13\pm 0.06\,.
\end{equation}
This correction factor is largely independent of the values of the Wilson coefficients, which cancel in the ratio, hence it can be applied both in the SM and in a wide class of SM extensions.\footnote{We refer here to the motivated class of SM extensions where non-standard contributions do not introduce sizable new local operators different from those present in (\ref{eq:bsll}).} 
In principle, multi-body modes such as $B\to K\pi\pi \llpair$ can also contribute to the total inclusive rate. However, these modes are suppressed even further by phase space factors and are expected to give a correction well within the current 
uncertainties of one- and two-body modes.

Using $\Delta^{[15]}_{K\pi}$, the semi-inclusive 
branching fraction  obtained summing over one- and two-body modes can be written as 
\begin{equation}
\begin{split}
    \sum_i&\mathcal{B}(B\to X^i_s\llpair)_{[15]} = \big(1 + \Delta_{K\pi}^{[15]}\big)\\ 
    &\times \big[\mathcal{B}(B\to K\llpair)_{[15]} + \mathcal{B}(B\to K^*\llpair)_{[15]}\big]\,,
    \label{eq:BRsinc1}
\end{split}
\end{equation}
both within and beyond the SM.
Combining (\ref{eq:BRsinc1}) and \eqref{eq:leadingmodes}, we 
arrive at the following SM estimate of the semi-inclusive 
branching fraction ($q^2\ge 15$ GeV):
\begin{equation}\label{eq:BRsemincl}
    \sum_i\mathcal{B}(B\to X^i_s\llpair)^{\rm SM}_{[15]} = \big(5.07\pm 0.42\big)\times 10^{-7}\,.
\end{equation}
As can be seen, this result is well-compatible with the truly inclusive estimate 
presented in Eq.~(\ref{eq:BRincl}). The compatibility of these two results can be viewed both as a consistency check of the form factor calculations~\cite{Parrott:2022rgu,Horgan:2015vla} or, alternatively, as a 
consistency check of the inclusive result in Eq.~(\ref{eq:BRincl}).

\subsection{Comparison with data}

\begin{figure}[t]
    \centering
    \includegraphics[width=0.65\linewidth]{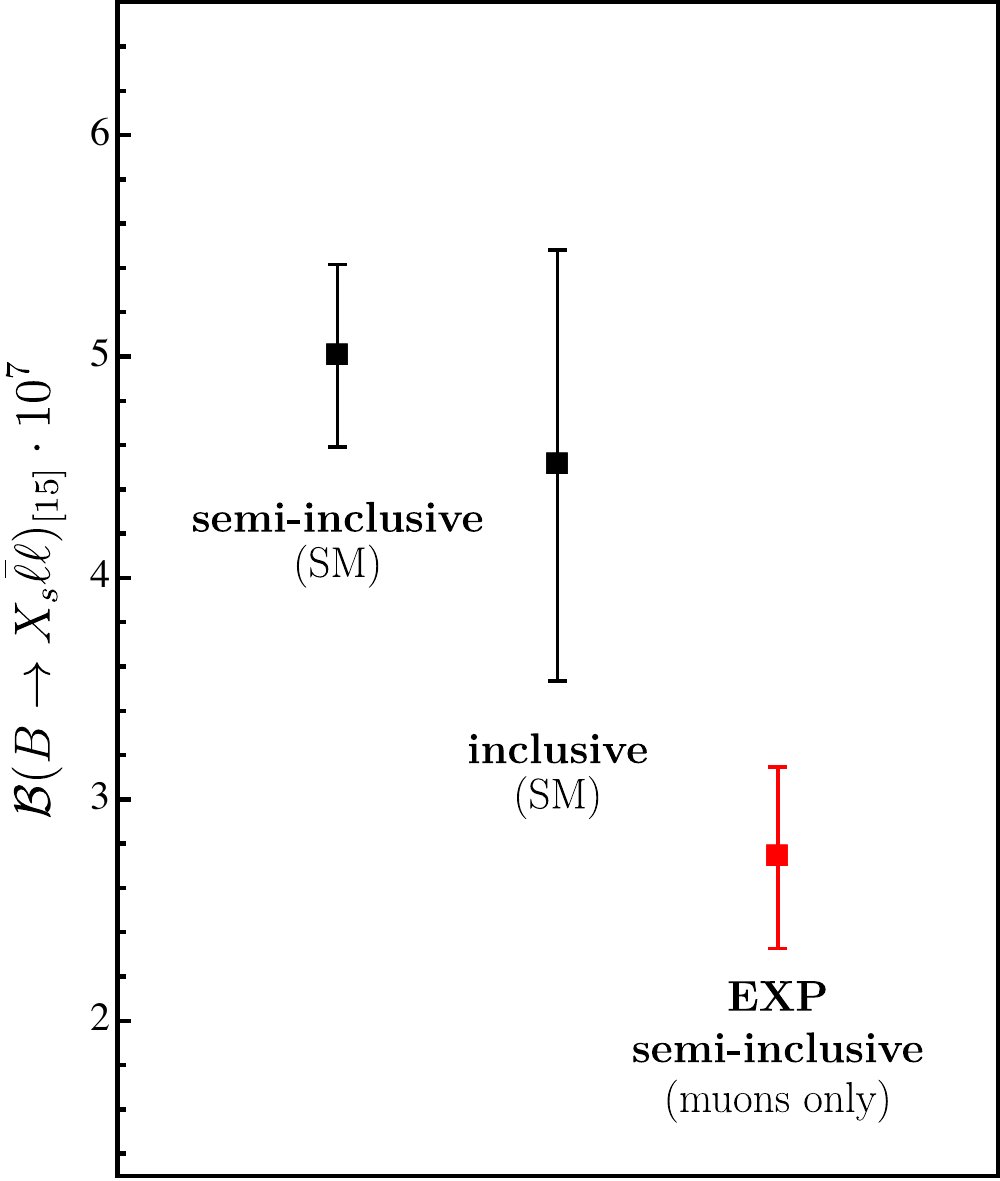}
    \caption{SM predictions vs.~experimental data  for the inclusive 
    branching ratio,  $\cB(B \to X_s \llpair)$, in the region $q^2 \geq 15~{\rm GeV}^2$.
    }
    \label{fig:BR}
\end{figure}

The experimental determinations of the two leading modes in the high-$q^2$ region,
and in the $\ell=\mu$ case, 
can be extracted from the results of the LHCb collaboration, Ref.~\cite{LHCb:2014cxe}
\begin{equation}
    \begin{split}
        \mathcal{B}(B\to K\mmpair)^{\rm exp}_{[15]} =& (8.47\pm 0.50)\times 10^{-8}\,, \\
        \mathcal{B}(B\to K^*\mmpair)^{\rm exp}_{[15]}  =& (1.58\pm 0.35)\times 10^{-7}\,.
    \end{split}
\end{equation}
Applying the correction factor in Eq.~\eqref{eq:correction}, 
we determine the following result for the measured semi-inclusive branching fraction:
\begin{equation}
    \sum_i\mathcal{B}(B\to X^i_s\mmpair)^{\rm exp}_{[15]}  = (2.74\pm 0.41) \times 10^{-7}\,.
    \label{eq:BRexp}
\end{equation}
As summarised in Fig.~\ref{fig:BR}, this result is significantly below the 
(consistent) SM predictions in  Eqs.~(\ref{eq:BRincl})
and~(\ref{eq:BRsemincl}).

\section{Discussion}
\label{sect:V}

The difference between the experimental result in (\ref{eq:BRexp})
and the SM predictions in  Eqs.~(\ref{eq:BRincl}) and~(\ref{eq:BRsemincl})
confirms the finding of several groups of a sizable suppression 
of the observed $b\to s\mmpair$ rates compared to SM expectations (see e.g.~\cite{Gubernari:2022hxn,Alguero:2023jeh,Altmannshofer:2021qrr} for recent analyses). The novel aspect of our analysis is that we 
support this conclusion, despite with a lower significance, 
by means of the inclusive rate at high $q^2$
in (\ref{eq:BRincl}), which is {\em insensitive} 
to hadronic form factors. 
We thus provide an important independent verification of this phenomenon.

The inclusive rate in the high-$q^2$ region also has a different sensitivity to non-perturbative effects associated with charm re-scattering, compared to exclusive observables (rates and angular distributions) in the low-$q^2$ region. We stress this point given that 
non-perturbative effects induced by charm re-scattering have been invoked as a possible SM explanation for the (lepton-universal) anomalies observed in the low-$q^2$ region~\cite{Ciuchini:2022wbq}.

\begin{figure}[t]
    \centering
    \includegraphics[width=0.95\linewidth]{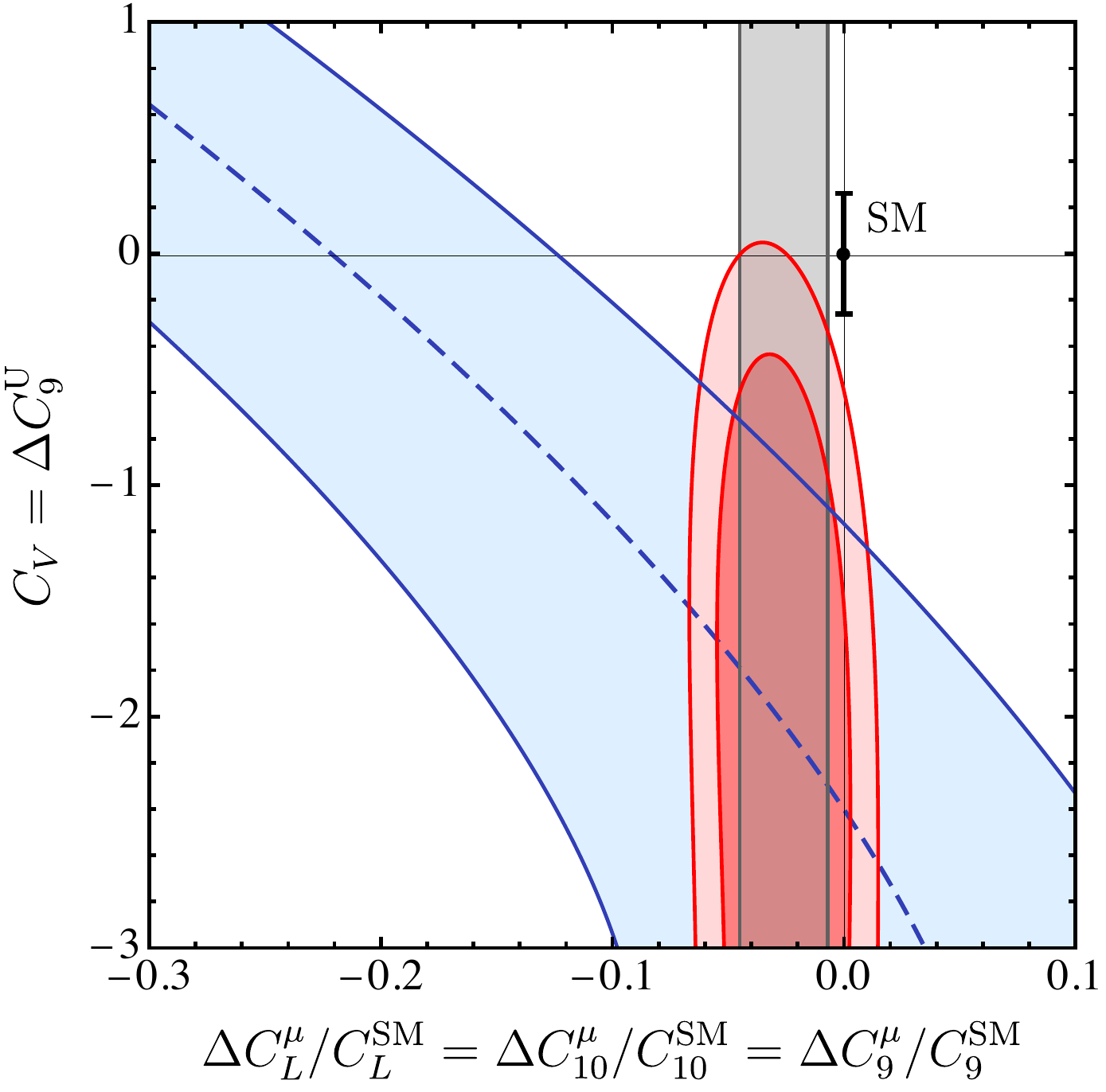}
    \caption{Regions for the Wilson coefficients favored by experimental data.
    Here, $\Delta C^\mu_L = C^\mu_L - C_L^{\text{SM}}$ is the correction to the SM value of $C_L$ for the muon modes. The blue area is the   $1\sigma$ compatibility region between the inclusive computation of $\mathcal{B}(B\to X_s\llpair)$ and the experimental sum of exclusive modes (the dashed line indicates the best fit). The vertical grey band shows the  $1\sigma$ value of 
    $C^\mu_L$ determined by  
    $\cB(B_s\to \mmpair)$ and the LFU ratios (assuming a lepton-universal $C_V$). The dark and light red regions give the combined compatibility at 68\% and 90\% confidence level, respectively. To ease the comparison with previous studies, we also show on both axes the notation in the standard operator basis.}
    \label{fig:CVCL}
\end{figure}

The fact that the observed discrepancy is hardly explained by charm re-scattering, especially for the inclusive rate, can be better appreciated by looking at the size of the effect in the operator basis defined in Section~\ref{sec:Lagrangian}.
In Fig.~\ref{fig:CVCL} we plot the region in the $C_V$--$C_L$ plane favored by present data, i.e.~treating $C_V$ and $C_L$ in Eq.~(\ref{eq:RinclCVCL}) as free parameters and fitting the experimental result in (\ref{eq:BRexp}). 
As already discussed in Section~\ref{sec:charmloops},
 both perturbative and non-perturbative contributions 
due to charm re-scattering  can be accounted for via an effective ($q^2$-dependent) modification
of $C_V$.
Assuming $C_L=C_L^{\rm SM}$, the modification of $C_V$ necessary to describe the data is very large: 
it is 
larger, and opposite in sign, with respect to 
the perturbative estimate of 
charm re-scattering 
contributions leading to Eq.~(\ref{eq:CVeff1}).

The central value of the discrepancy 
is beyond any realistic estimate of non-perturbative charm re-scattering far from the narrow-charmonium region. The latter are not enhanced by RG logarithms, 
and in the high-$q^2$ region are expected to be of 
$O(\Lambda^2_{\rm QCD}/q^2)$. Explicit estimates 
of these effects for the  
inclusive rate~\cite{Huber:2019iqf} and the 
leading exclusive modes~\cite{Beylich:2011aq} lead to modifications of $C_V$ of  $O(1\%)$. 
Even assuming an order of magnitude enhancement (which would be hard to justify~\cite{Beylich:2011aq,Huber:2019iqf}), these contributions are within 
the error band shown in Fig.~\ref{fig:CVCL} (SM point), that we deduce from the scale variation of the perturbative contribution (as
already stated, 
we assume the scale variation 
of $C^{\rm eff}_V(q^2)|^{(0)}_{\rm pert}$ in (\ref{eq:CVeff1})
as uncertainty for the SM estimate of $C_V$).

In exclusive modes, and specific values of $q^2$,
large violations of quark-hadron duality are certainly possible. For instance, the large violations of 
naive factorization observed in~\cite{Lyon:2014hpa}, 
also in the 
high-$q^2$ region, are a manifestation of this statement.
However, we stress that we are considering an inclusive quantity, where such effects are expected to be much smaller~\cite{Beneke:2009az}.
In conclusion, although we are unable to provide a rigorous upper 
bound on charm re-scattering contributions,
we believe that the size of 
$C^{\rm eff}_V(q^2)|^{(0)}_{\rm pert}$  and the explicit estimates of
non-perturbative effects 
presented in~\cite{Beneke:2009az,Beylich:2011aq}, 
indicate that such effects  
cannot account for the bulk of the difference between SM and experimental points in Fig.~\ref{fig:BR}.

In Fig.~\ref{fig:CVCL} we also show the impact of a possible change in 
$C_L$, which can occur only beyond the SM. More precisely, we consider the motivated case (see e.g.~\cite{Alguero:2018nvb,Cornella:2021sby}) of a lepton non-universal modification of $C_L$, 
affecting the muon modes only, versus a lepton-universal shift in $C_V$.\footnote{The lepton universal nature of $C_V$ is a natural consequence of its vector-like structure: this effective operator can appear naturally (i.e.~without a tuning between left-handed and right-handed components) via 
an effective short-distance interaction of the type $(\overline{s}_L\gamma_{\mu} b_L) D_\nu F^{\mu\nu}$, which is necessarily lepton universal.}
The value of  $\Delta C_L^\mu$ is strongly constrained by 
$\cB(B_s\to \mmpair)$ and the Lepton Flavor Universality (LFU) ratios 
($R_K$ and $R_{K^*}$).
Updating the analysis of Ref.~\cite{Cornella:2021sby}
taking into account the superseded values 
of the LFU ratios in~\cite{LHCb:2022qnv,LHCb:2022zom}, together with the $R_{K_S}$ and $R_{K^{*+}}$ results in~\cite{LHCb:2021lvy}, 
and adding to the $\cB(B_s\to \mmpair)$ data the CMS result in~\cite{CMS:2022mgd}, 
we find $\Delta C_L^\mu = -0.026 \pm 0.019$. 
This result, taken alone, does not indicate a significant deviation from the SM; however, combining it 
with the constraint from $\mathcal{B}(B\to X_s\mmpair)_{[15]}$ leads to a preferred region in the $C_V$--$\Delta C^\mu_L$ plane 
which does not include the SM point at the $90\%$~C.L.

On general grounds, beyond-the-SM contributions to the Wilson coefficients are expected to be small corrections over the SM ones (evaluated at the electroweak scale). Fig.~\ref{fig:CVCL} shows that this condition cannot be satisfied if we assume non-standard contributions to $C_V$ only. On the other hand, this condition can be satisfied for both $C_V$ and $\Delta C^\mu_L$, but only  if $|\Delta C^\mu_L| \not =0$, 
hence in the presence of a small but non-negligible LFU-violating amplitude.

\section{Conclusions}

The inclusive  $B \to X_s\llpair$ rate at high dilepton invariant mass
provides a clean and sensitive probe of $b\to s\llpair$ amplitudes.
In this paper we have presented an updated estimate of this rate within the SM, by means of the ratio~(\ref{eq:R0}) and Belle's data on $\Gamma(B \to X_u \bar\ell\nu)$~\cite{Belle:2021ymg}.
The result, shown in Fig.~\ref{fig:BR}, is in good agreement with the semi-inclusive estimate obtained summing the leading one-body modes ($K$ and $K^*$) and the subleading non-resonant $K\pi$ channel 
in the relevant kinematical region. The uncertainty on the fully inclusive prediction is sizable, but it is dominated by the experimental error on $\Gamma(B \to X_u \bar\ell\nu)$, hence it could be significantly improved in the near future. 
 
The good compatibility between inclusive and semi-inclusive SM predictions confirms the expectation that the inclusive rate, in the kinematical region
$q^2~\geq~15~{\rm GeV}^2$, is dominated by few exclusive modes. 
This opens up the possibility of a precise comparison with data for 
the rare decays with muon modes collected by LHCb.
This comparison, also shown in Fig.~\ref{fig:BR}, 
confirms the finding of several groups of a sizable suppression of the observed $b\to s\mmpair$ rates compared to SM expectation.
The evidence of this effect from the inclusive high-$q^2$ rate does not exceed $2\sigma$, but it is not based on hadronic form factors hence providing an important independent verification of this phenomenon.

As a byproduct of our analysis, we have shown that it is more convenient to describe short-distance contributions to $b\to s\llpair$ amplitudes, both within and beyond the SM, via the effective operators $\cO_V$--$\cO_L$ in Eq.~(\ref{eq:OVOL}), rather than in the standard $\cO_9$--$\cO_{10}$ basis. 
In this basis, BSM contributions to  
$\cO_V$ are naturally lepton universal, 
whereas those to $\cO_L$ can be lepton-flavor 
dependent. The best-fit values 
in the $C_V$--$C^\mu_L$ plane, following from the present analysis, combined with recent data on LFU ratios and $\cB(B_s\to \mmpair)$, are shown in Fig.~\ref{fig:CVCL}. This analysis indicates that explaining the bulk of the present discrepancy via a modification to $C_V$ only would require an 
unnaturally large correction to $C_V$, which we cannot justify via underestimated non-perturbative effects
(and is also unlikely to appear in realistic BSM theories). 
 By contrast, relatively small BSM effects to both $C_V$ and $C^\mu_L$ can describe the data well.

As already stressed, uncertainties at present are still large, but the theory errors play a subleading role. We thus expect that the method outlined in this paper can have a significant impact in the near future in shedding light on the interesting puzzle of $b\to s\llpair$ transitions. 

\subsection*{Acknowledgements}

This project has received funding from the European Research Council~(ERC) under the European Union's Horizon~2020 research and innovation programme under grant agreement 833280~(FLAY), and by the Swiss National Science Foundation~(SNF) under contract~200020\_204428.

\bibliography{references}
\end{document}